\documentclass[aps, prl, twocolumn, superscriptaddress]{revtex4}
\usepackage{graphicx}
\usepackage{dcolumn}
\usepackage{bm}
\usepackage{natbib}
\usepackage{color}

\begin{document}

\title{Terahertz Quantum Hall Effect in a Topological Insulator }

\author{A.~M.~Shuvaev}
\affiliation{Institute of Solid State Physics,
Vienna University of Technology, 1040 Vienna, Austria}%
\author{G.~V.~Astakhov}
\affiliation{Physikalisches Institut (EP6), Universit\"{a}t
W\"{u}rzburg, 97074 W\"{u}rzburg, Germany}%
\author{G.~Tkachov}
\affiliation{Institut f\"ur Theoretische Physik und Astronomie, Universit\"{a}t
W\"{u}rzburg, 97074 W\"{u}rzburg, Germany}%
\author{C.~Br\"{u}ne}
\author{H.~Buhmann}
\author{L.~W.~Molenkamp}
\affiliation{Physikalisches Institut (EP3), Universit\"{a}t
W\"{u}rzburg, 97074 W\"{u}rzburg, Germany}%
\author{A.~Pimenov}
\affiliation{Institute of Solid State Physics,
Vienna University of Technology, 1040 Vienna, Austria}%

\date{\today}

\begin{abstract}

Using THz spectroscopy in external magnetic fields we investigate
the low-temperature charge dynamics of strained HgTe, a three
dimensional topological insulator. From the Faraday rotation angle
and ellipticity a complete characterization of the charge carriers
is obtained, including the 2D density, the scattering rate and the
Fermi velocity. The obtained value of the Fermi velocity provides
further evidence for the Dirac character of the carriers in the
sample. In resonator experiments, we observe quantum Hall
oscillations at THz frequencies. The 2D density estimated from the
period of these oscillations agrees well with direct transport
experiments on the topological surface state. Our findings open new
avenues for the studies of the finite-frequency quantum Hall effect
in topological insulators.

\end{abstract}


\maketitle

Three dimensional topological insulators
\cite{hasan_rmp_2010,qi_prb_2008} have attracted much interest
recently, as they exhibit  a number of unusual and non-trivial
properties, such as protected conducting states on the surfaces of the sample.
Unusual electrodynamics, such as a
universal Faraday effect and an anomalous Kerr rotation have been
predicted
\cite{tse_prl_2010,tse_prb_2011,maciejko_prl_2010,tkachov_prb_2011}
for these surface states, their observation is still
outstanding.  We  showed recently that strained HgTe, where the strain
lifts the light-hole--heavy-hole degeneracy
that normally is present in bulk HgTe, is a very
promising 3D topological insulator \cite{brune_prl_2011}. This is
because at low temperatures parasitic effects due to bulk carriers are
practically absent.
In static transport experiments a strained 70 nm thick HgTe layer
\cite{brune_prl_2011} exhibits a quantum Hall effect (QHE),
yielding direct evidence that the charge carriers
in these layers are confined to the topological two dimensional (2D)
surface states of the material. These findings are further corroborated
by recent Faraday rotation data \cite{hancock_prl_2011} in a similar layer,
which have been obtained using a terahertz time-domain technique.

In this work, we present the results of low temperature terahertz
Faraday cw transmission experiments on another strained HgTe film.
The carrier density, Fermi velocity and the scattering rate can be
reliably determined from these data. In particular, we obtain the
Fermi velocity $v_F = 0.52 \cdot 10^6$ m/s, which is in excellent
agreement with the Faraday rotation experiments
\cite{hancock_prl_2011} and the dc Shubnikov-de Haas measurements
\cite{brune_prl_2011} on 70-nm-thick strained HgTe films as well as
with band-structure calculations for the surface states in 3D
topological insulators (see e.g. Ref. \cite{liu_prb_2010}). In the
same sample we observe quantum Hall-induced oscillations at
terahertz frequencies, providing further evidence for the 2D
character of the conductivity. In the case of topological insulators,
no finite frequency QHE has been reported up to now.

The sample studied in this work is a coherently strained 52-nm-thick
nominally undoped HgTe layer, grown by molecular beam epitaxy on an
insulating CdTe substrate \cite{becker_pss_2007}. Transmittance
experiments at terahertz frequencies (100 GHz $< \nu <$ 800 GHz)
have been carried out in a Mach-Zehnder interferometer
arrangement~\cite{volkov_infrared_1985, pimenov_prb_2005} which
allows measurement of the amplitude and phase shift of the
electromagnetic radiation in a geometry with controlled
polarization. Using wire grid polarizers, the complex transmission
coefficient can be measured both in parallel and crossed polarizers
geometry. Static magnetic fields, up to 8~Tesla, have been applied
to the sample using a split-coil superconducting magnet.

To interpret the experimental data we use the ac conductivity
tensor $\hat{\sigma} (\omega)$ obtained in the classical (Drude)
limit from the Kubo conductivity of topological surface states (see
e.g. Ref. \cite{tse_prb_2011}). The diagonal, $\sigma_{xx}
(\omega)$, and Hall, $\sigma_{xy} (\omega)$, components of the
conductivity tensor as functions of THz frequency $\omega$ can be
written as:
\begin{eqnarray}
&& \sigma_{xx} (\omega)=\sigma_{yy} (\omega) =
\frac{1-i \omega \tau}{(1-i \omega \tau)^2 +(\Omega_c \tau)^2} \sigma_0
\,, \label{sxx}\\
&& \sigma_{xy} (\omega)=-\sigma_{yx} (\omega)= \frac{\Omega_c
\tau}{(1-i \omega \tau)^2 +(\Omega_c \tau)^2} \sigma_0 \,.
\label{sxy}
\end{eqnarray}
Here, $\Omega_c = eBv_F/\hbar k_F$ is the cyclotron frequency, $\sigma_0$ is
the dc conductivity, $B$ is the magnetic field, $v_F$, $k_F$, $e$, and
$\tau$ are the Fermi velocity, Fermi wave-number, charge, and scattering time of
the carriers, respectively. For the Dirac spin-helical surface states the Fermi
wave-number depends
on the 2D carrier density, $n_{2D}$, through relation $k_F=\sqrt{4\pi n_{2D}}$,
with no spin degeneracy.

The transmission spectra can then be calculated using a transfer matrix formalism
\cite{berreman_josa_1972,shuvaev_epjb_2011,shuvaev_prl_2011} which
takes multiple reflection within the substrate into account. The
electrodynamic properties of the CdTe substrate have been obtained
in a separate experiment on a bare substrate. Further details of the
fitting procedure
can be found in the Supplementary information to Ref.
\cite{shuvaev_prl_2011}. Neglecting any substrate effects,
the complex transmission coefficients in parallel
($t_p$) and crossed ($t_c$) polarizers geometry can be written as:
\begin{eqnarray}
&& t_p =\frac{4+2\Sigma_{xx}}
{4+4\Sigma_{xx}+\Sigma_{xx}^2+\Sigma_{xy}^2}  \,, \label{tp}\\
&& t_c =\frac{2\Sigma_{xy}}
{4+4\Sigma_{xx}+\Sigma_{xx}^2+\Sigma_{xy}^2}  \,. \label{tc}
\end{eqnarray}
Here $\Sigma_{xx}$ and $\Sigma_{xy}$ are effective dimensionless 2D
conductivities, defined as: $\Sigma_{xx}=\sigma_{xx}dZ_0$ and
$\Sigma_{xy}=\sigma_{xy}dZ_0$ with the HgTe film thickness
$d=52$\,nm and the vacuum impedance $Z_0 \approx 377\,\Omega$. In
order to self-consistently obtain the parameters of the
quasiparticles, the  field-dependent complex transmission $t_p(B)$
and $t_c(B)$ for $\nu =$0.17 THz, 0.35 THz and 0.75 THz and the
zero-field transmittance spectra $|t_c(\omega)|^2$ have been fitted
simultaneously.

\begin{figure}
\includegraphics[width=0.6\linewidth, clip]{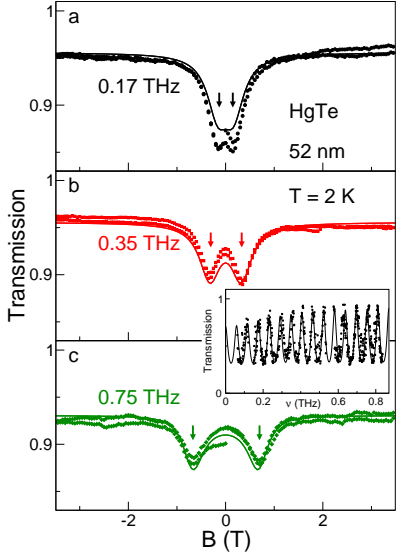}
\caption{\emph{Magnetic field dependence of the transmission in strained
HgTe.} (a-c) Transmission amplitude in parallel polarizers ($t_p$)
geometry, showing cyclotron resonance at the positions indicated by the arrows.
The frequency of the experiments is indicated in the
panels. The inset shows the frequency dependent transmittance in
zero external magnetic field, $|t_p(B=0)|^2$. Symbols: experiment,
solid lines: simultaneous fit of all data with the Drude model as
described in the text.} \label{ftran}
\end{figure}

The inset in Fig. \ref{ftran} shows the transmittance spectrum of the
HgTe film at zero magnetic field. The characteristic
oscillations in the spectrum, with a period of about 58 GHz, are due to
Fabry-P\'{e}rot type interferences within the CdTe substrate. The
absolute transmittance in the interference maxima is close to 95\%,
which reflects the low effective conductance of our HgTe film,
$\Sigma_{xx} \ll 1$. At low frequencies, the maximum transmittance
decreases and approaches $|t_p|^2 \simeq 0.7$ in the zero
frequency limit. Such a behavior is typical for Drude carriers
with a scattering rate in the frequency region of the experiment.
Indeed, the solid line in the transmission spectra represents a
Drude fit with the parameters given in the first row of Tab. \ref{tab}.

From the fits we obtain the Fermi velocity $v_F = 0.52 \cdot 10^6$
m/s. This value is very close both to $v_F = (0.51 \div 0.58) \cdot
10^6$ m/s as determined in the Faraday rotation experiments on a
70-nm-thick strained HgTe film \cite{hancock_prl_2011} and to $v_F =
0.42 \cdot 10^6$ m/s as extracted from dc Shubnikov-de Haas
measurements on a patterned 70-nm-thick strained HgTe layer
\cite{brune_prl_2011}. The obtained value of the Fermi velocity is
also in very good agreement with the band-structure-theory result
$v_F = 0.51 \cdot 10^6$ m/s for the linear (Dirac) part of the
surface-state spectrum in topological insulators (see e.g. Ref.
\cite{liu_prb_2010}). As an additional check of the 2D surface
carrier dynamics in our sample, we have analyzed the terahertz
transmission data of Ref. \cite{shuvaev_prl_2011} for a 70-nm-thick
strained HgTe film at high temperature $T=200$ K and for a bulk
(1000-nm-thick) unstrained HgTe sample. In both cases, the
electrodynamics is governed by massive bulk carriers, for which the
values of $v_F$ turn out to be much larger than the Dirac
surface-state velocity, i.e.,  $v_F \approx 0.5 \cdot 10^6$ m/s
(Tab. \ref{tab}).

\begin{table*}
\caption{Drude parameters of the charge carriers in HgTe in strained
and unstrained films. The data on 70 nm and 1000 nm film were partly
given in Ref. \cite{shuvaev_prl_2011}. } \label{tab}
\begin{ruledtabular}
\begin{tabular}{lrrrrr}
Sample & $T$(K) & $n_{2D}$(cm$^{-2}$) & $v_F$(ms$^{-1}$)  & $1/2\pi
\tau$
(GHz) & $G_{2D}=\sigma_0 \cdot d\ (\Omega^{-1})$ \\
\hline \hline
52 nm (strained) [this work] & 2 & $1.08\cdot10^{11}$ & $0.52\cdot 10^6$
& 250 & $7.6\cdot 10^{-4}$ \\
\hline
70 nm (strained) \cite{shuvaev_prl_2011} & 4   & $4.8\cdot10^{10}$
& $0.38\cdot 10^6$  & 210 & $4.3\cdot 10^{-4}$ \\
& 200 & $1.5\cdot10^{12}$ & $1.63\cdot 10^6$   & 360 & $5.3\cdot 10^{-3}$ \\
\hline
1000 nm (unstrained) \cite{shuvaev_prl_2011} & 3 & $4.2\cdot10^{11}$
& $0.99\cdot 10^6$  & 240 & $2.8\cdot 10^{-3}$ \\
& 200 & $4.9\cdot10^{13}$ & $9.36\cdot 10^6$  & 360 & $1.9\cdot 10^{-1}$\\
\end{tabular}
\end{ruledtabular}
\end{table*}


\begin{figure}
\includegraphics[width=0.9\linewidth, clip]{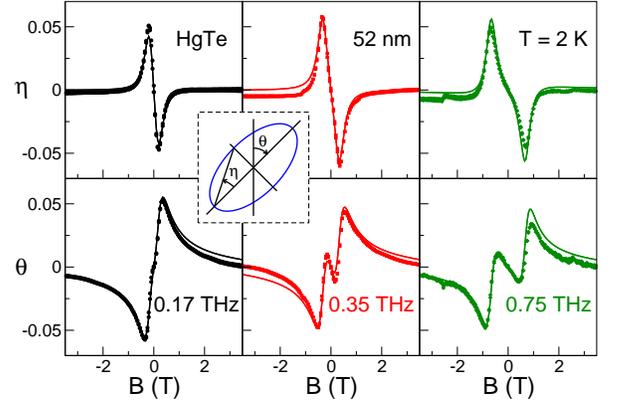}
\caption{\emph{Complex Faraday angle $\theta + i \eta$ in HgTe.}
Bottom panels: Faraday rotation, top panels: ellipticity for the
same frequencies as in Fig. \ref{ftran}. The inset sketches the
definitions of the Faraday rotation $\theta$ and ellipticity $\eta$.
Symbols: experiment, solid lines: simultaneous fit of all data with
the Drude model as described in the text. Angular units are
radians.} \label{fang}
\end{figure}

Figure \ref{ftran} shows the magnetic field dependent transmittance of the HgTe
film in Faraday geometry and for parallel orientation of
polarizer and analyzer. According to Eq. (\ref{tp}), the
transmittance in parallel polarizers ($t_p$) depends mainly on
$\Sigma_{xx}$. For all three frequencies two clear minima in the
transmitted signal are observed in the range below $\pm 1$\,T. The minima
in $|t_p|$ roughly correspond to the cyclotron resonance energy and scale
with magnetic field. This may be
understood taking into account that in our case $\Sigma \ll 1$ and
Eqs. (\ref{tp},\ref{tc}) simplify to:
\begin{equation}\label{trsimple}
    t_p \simeq 1- \Sigma_{xx}/2; \quad t_c \simeq \Sigma_{xy}/2 \ .
\end{equation}
In the limit $\omega\tau \gg 1$, Eq. (\ref{sxx}) may be approximated by
\begin{equation}\label{sxx1}
    \sigma_{xx} \simeq \frac{1-i \omega \tau}{(\Omega_c ^2-\omega ^2)\tau^2}
\sigma_0  \ ,
\end{equation}
which leads to a resonance like feature for $\Omega_c=\omega$.
Thus, the positions and widths of the minima in Fig.
\ref{ftran} are directly connected with the
parameter $v_F/k_F$ and the scattering rate $\tau^{-1}$ of the charge carriers.

Figure \ref{fang} shows the complex Faraday angle $\theta + i \eta$
as obtained at the same frequencies as in Fig. \ref{ftran}. The
polarization rotation $\theta$ and the ellipticity $\eta$ are
obtained from the transmission data using:
\begin{eqnarray}
&& \tan(2\theta)=2\Re(\chi)/(1-|\chi|^2)\ , \\
&& \sin(2\eta)=2\Im(\chi)/(1+|\chi|^2)\ .
\end{eqnarray}
Here $\chi=t_c/t_p$ and the definitions of $\theta + i
\eta$ are shown graphically in the inset to Fig. \ref{fang}. A direct
interpretation of the complex Faraday angle is in general not
possible because of the interplay of $\sigma_{xx}$ and $\sigma_{xy}$
in the data.

In the low frequency limit, $\omega\tau \ll 1$  Eq. (\ref{sxy})
simplifies to the static result $\sigma_{xy} = \Omega_c \tau\sigma_0
/(1 +(\Omega_c \tau)^2) $. The last expression has a maximum at
$\Omega_c (B) = \tau^{-1} $, which leads to maxima in $t_c$ and $\theta$
at about the same field value. Therefore, the Faraday angle provides
a direct and an independent way of obtaining the scattering rate
$1/\tau$. The solid line in Fig. \ref{fang} are the fits which have
been done simultaneously for all results presented above. In total,
the parameters of the charge carriers have been obtained by
simultaneously fitting ten data sets. The quite reasonable fit of
all results proves that a single type of charge carriers dominates
the electrodynamics in the range of frequencies and magnetic fields
used in these experiments.

\begin{figure}
\includegraphics[width=0.7\linewidth, clip]{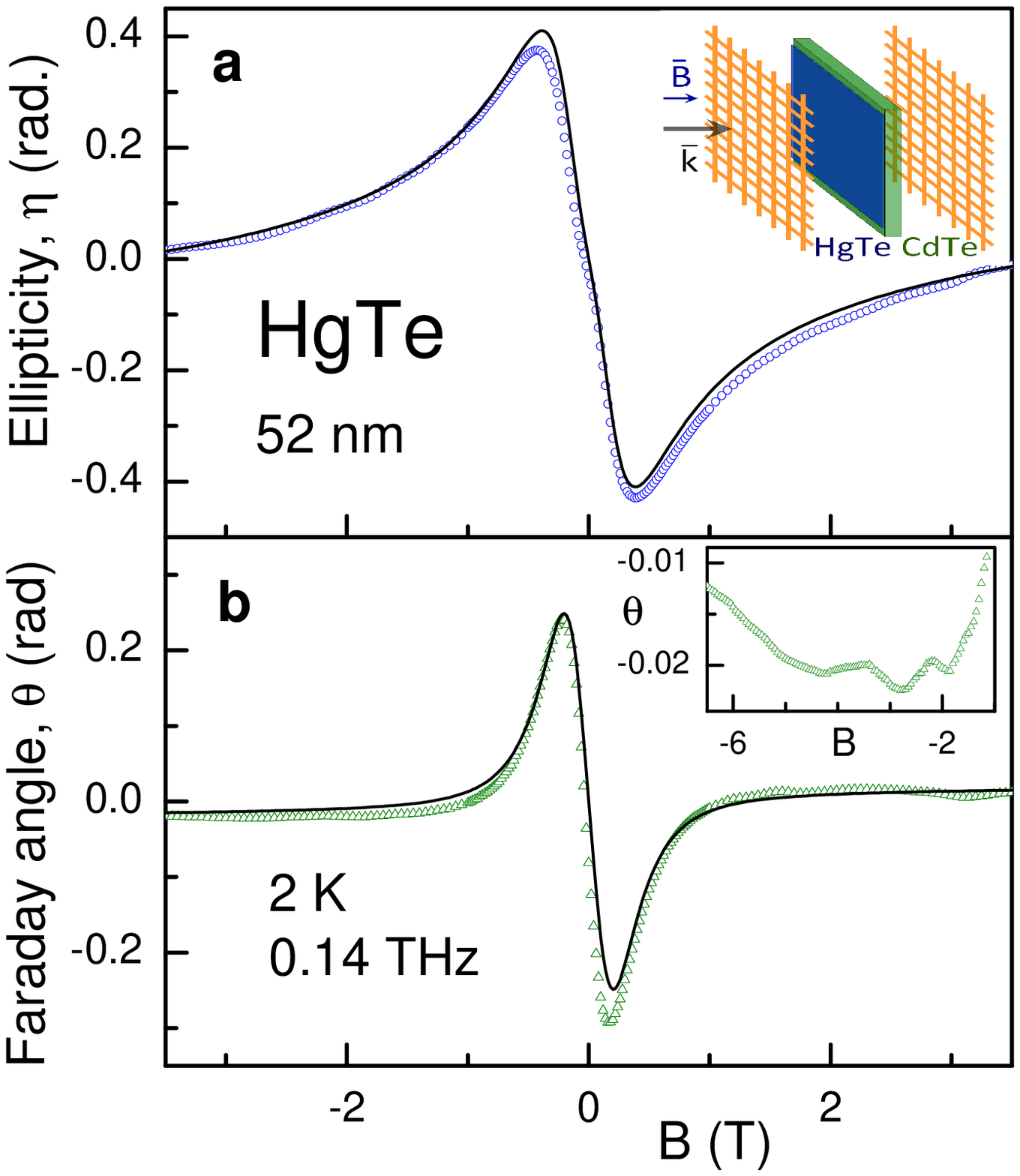}
\caption{\emph{Faraday rotation in HgTe within resonator geometry.}
(a) - Ellipticity, (b) - Faraday angle. Symbols - experiment, lines
- fits according to Eqs. (1-4). Upper inset shows the experimental
geometry within a Copper meshes resonator. Lower inset shows a
magnified view of the Faraday angle demonstrating QHE oscillations.}
\label{fres}
\end{figure}

Very solid evidence for the two dimensional character of the
carriers probed in the Faraday rotation experiments would be the
observation of quantum Hall plateaus, similar to the observation of
the QHE in \cite{brune_prl_2011}. However, the accuracy of the
experiments shown above does not allow to observe the QHE. In order
to solve this problem, we have performed further Faraday
transmission experiments on the same sample, now using a resonator
geometry as shown in the inset of Fig. \ref{fres}.

\begin{figure}
\includegraphics[width=0.95\linewidth, clip]{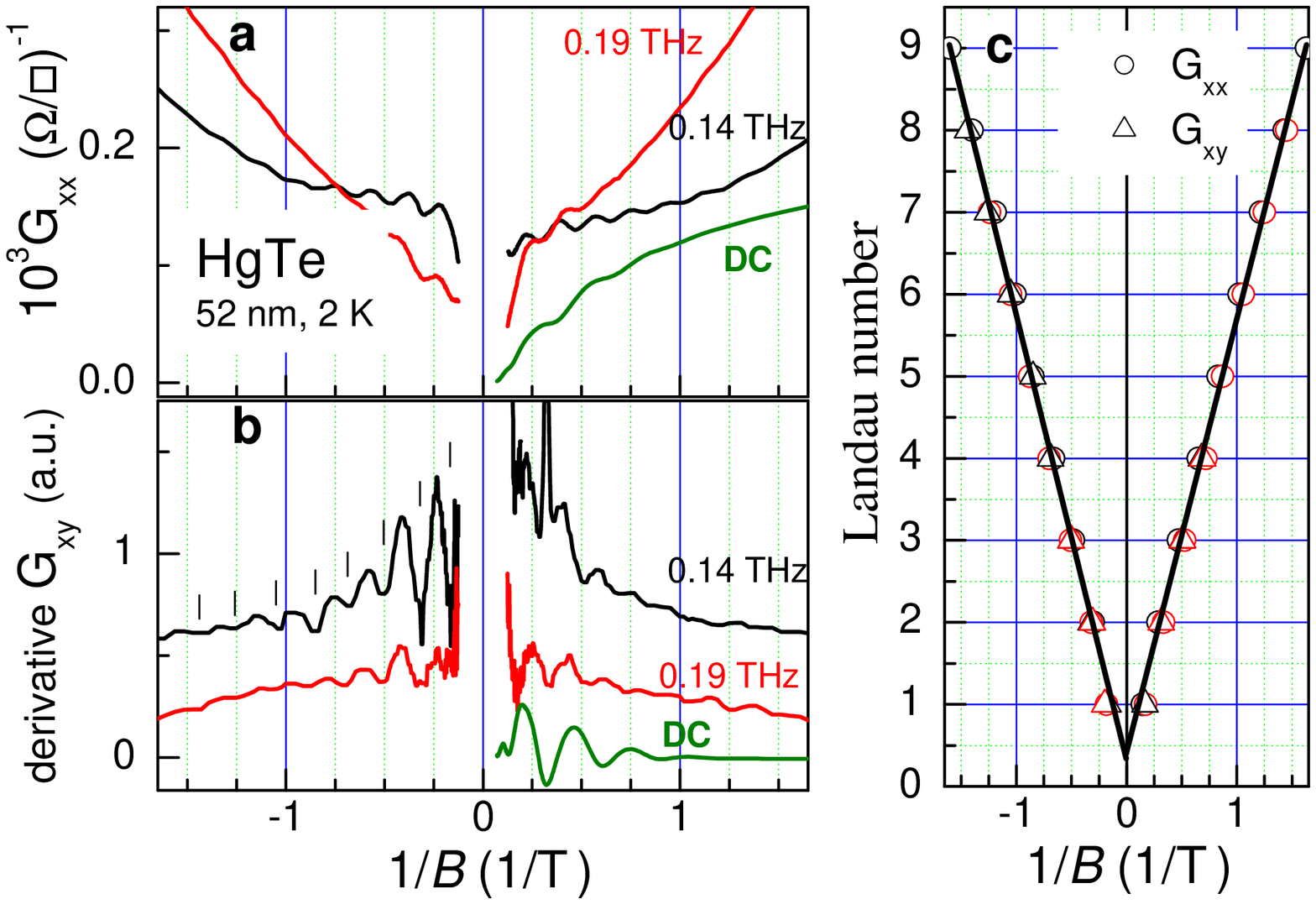}
\caption{\emph{Terahertz quantum Hall effect in HgTe.} (a) Two
dimensional conductance: $G_{xx}$, (b) derivative of $G_{xy}$
($dG_{xy}/dB^{-1}$). The data have been obtained within a resonator
geometry and are plotted as a function of inverse magnetic field.
The experimental data are shown as solid lines for frequencies as
indicated. Dashes in the bottom panel marks the minima for negative
magnetic fields. (c) Numbered positions of the minima in $G_{xx}$
and in the derivative of $G_{xy}$ for 0.14 THz and 0.19 THz.
Straight lines yield interpolation to the origin. } \label{fqhe}
\end{figure}

In these experiments, the sample is placed in the middle of a
Fabry-P\'{e}rot resonator defined by metallic meshes. We have
utilized Cu meshes with a 200 $\mu$m period. The
distance between adjacent maxima of the resonator is $\simeq
51$\,GHz. In the frequency range between 100 and 200 GHz the quality
factor of the loaded resonator is about $Q \sim 10$. This indicates
that, effectively, the radiation passes about ten times through the
sample before reaching the detector, which effectively increases the
sensitivity to fine details by roughly the same value. As shown in
Fig. \ref{fres}, in the resonator experiments the field dependence
of the Faraday rotation and the ellipticity appears qualitatively similar
to that in Fig. \ref{fang}. An exact calculation of the complex
transmission coefficients within a resonator is complicated because
of the increased number of parameters. Therefore, in this case we
utilize the simple equations Eqs. (\ref{sxx})-(\ref{tc})
which neglect the effect of the substrate and the resonator completely.
Nevertheless, as clearly seen in Fig. \ref{fres}, the fits based on the
simplified expressions
reproduce the experimental results reasonably well. Fitting of
the signals for parallel and crossed polarizers yields within
experimental accuracy the same parameters as in the experiments
without a resonator. The only parameter which differs from the
results without a resonator is the absolute value of the
conductivity. This is of course expected, and results from multiple
transmission in the resonator and the influence of the substrate.

The main advantage of the resonator experiments is a higher
sensitivity to the details of the field-dependent transmission. In
addition to an overall field dependence similar to that in Figs.
\ref{ftran} and \ref{fang}, a tiny modulation of the signal can now be
observed. To convert this modulation to a conventional presentation,
we have inverted the transmittance curves into the 2D conductivity,
using Eqs. (\ref{tp} and \ref{tc}).
Because the absolute transmittance is not well-defined in the resonator
experiments,
we have scaled the absolute 2D conductance to agree with the data
without a resonator. The final results expressed in form of the
effective 2D conductance $G_{xx,xy}=\Sigma_{xx,xy}/Z_0$ are shown in
Fig. \ref{fqhe}.

Fig. \ref{fqhe}a shows the real part of the two dimensional
conductance $G_{xx}$ as a function of inverse magnetic field. Clear
oscillations in the conductance can be observed in this
presentation. In general, the phenomenology of the QHE at terahertz
frequencies is not well understood
\cite{hols_prl_2002,ikebe_prl_2010}. Existing experiments are
generally limited to frequencies below 100 GHz and they are analyzed
using scaling exponents \cite{sondi_rmp_1997,hols_prl_2002}. In the
resonator experiments, the field dependent oscillations can be
observed both with parallel and crossed polarizers. Contrary to
$G_{xx}$, the off-diagonal conductance $G_{xy}$ shows a substantial
field dependence even in high magnetic fields. Therefore, no clear
QHE signal can be directly detected in $G_{xy}$. In order to extract
the QHE information from these data, we have plotted the derivative
of the $G_{xy}$ as a function of an inverse magnetic field
($dG_{xy}/dB^{-1}$) in Fig. \ref{fqhe}b. The derivative has the
advantage of being insensitive to any residual slowly varying
signals, and, importantly, the expected steps in $G_{xy}$ are
transformed into the minima of the derivative. Finally, in order to
analyze the quantum Hall effect, both the minima in $G_{xx}$ and in
the derivative of $G_{xy}$ have been taken into account. In Fig.
\ref{fqhe}a,b the results at finite frequencies are compared with dc
QHE on the same sample. The periodicity of the oscillations in the
dc experiments is slightly different because of different carrier
concentration at the sample surface, induced by exposure to
photoresist and the presence of ohmic contacts.

The main results of the QHE experiments are represented in Fig.
\ref{fqhe}c demonstrating an approximate equidistant positioning of
all minima (labeled by number $N$) in inverse magnetic fields
$B^{-1}$ with the period of $\Delta B^{-1} = 0.18$ T$^{-1}$. This
periodicity reflects the dependence of the number of the occupied
Landau levels on $B^{-1}$. In a total, we detect the oscillations up
to index number $\pm 10$; also the first oscillations with the
Landau level index $\pm 1$ are clearly observed in the data. From
the periodicity of these oscillations the effective 2D carrier
density can be estimated according to the free electron expression
$n_{2D}=e/(h\Delta B^{-1}) \simeq 1.4 \cdot 10^{11}$cm$^{-2}$. This
value agrees reasonably well with the density $n_{2D}=1.08 \cdot
10^{11}$cm$^{-2}$ obtained directly from fitting the transmittance
and the Faraday rotation on the basis of the Drude model
(Tab.~\ref{tab}). Therefore, we may conclude that charge carriers
which are responsible for the terahertz electrodynamics at low
temperatures reveal 2D behavior. To further characterize the
electron system in our sample we extrapolated the dependence
$N(B^{-1})$ to the origin (see straight lines in Fig. \ref{fqhe}c),
which corresponds to the limit of very strong magnetic fields. At
the origin we find a finite value $N \approx 1/2$ instead of $N=0$
as would be the case for the conventional QHE. Previously, similar
extrapolated values were reported for graphene (see e.g. Ref.
\cite{novoselov_nature_2005}), zero-gap HgTe quantum wells
\cite{buttner_nphys_2011} and strained 70 nm-thick HgTe films
\cite{brune_prl_2011}, i.e. for materilas with 2D Dirac-like charge
carriers encoding a nonzero Berry phase. We therefore believe that
our teraherz QHE also indicates the 2D Dirac-like behavior.

In conclusion, we have analyzed the terahertz Faraday rotation in a
strained HgTe film. From these data all relevant parameters of the
charge carriers can be obtained. In addition, terahertz quantum
Hall effect oscillations have been observed within the same
experiment, which proved the two-dimensional character of the
conductivity.

We thank E. M. Hankiewicz for valuable discussion. This work was
supported by the by the German Research Foundation DFG (SPP 1285,
FOR 1162) the joint DFG-JST Forschergruppe on 'Topological
Electronics', the ERC-AG project '3-TOP', and the Austrian Science
Funds (I815-N16).

\bibliographystyle{unsrt}
\bibliography{lit_HgTe}

\begin{thebibliography}{10}

\bibitem{hasan_rmp_2010}
M.~Z. Hasan and C.~L. Kane.
\newblock \textit{Colloquium} : Topological insulators.
\newblock {\em Rev. Mod. Phys.}, 82:3045--3067, Nov 2010.

\bibitem{qi_prb_2008}
Xiao-Liang Qi, Taylor~L. Hughes, and Shou-Cheng Zhang.
\newblock Topological field theory of time-reversal invariant insulators.
\newblock {\em Phys. Rev. B}, 78:195424, Nov 2008.

\bibitem{tse_prl_2010}
Wang-Kong Tse and A.~H. MacDonald.
\newblock Giant magneto-optical kerr effect and universal faraday effect in
  thin-film topological insulators.
\newblock {\em Phys. Rev. Lett.}, 105:057401, Jul 2010.

\bibitem{tse_prb_2011}
W.-K. Tse and A.~H. MacDonald.
\newblock Magneto-optical faraday and kerr effects in topological insulator
  films and in other layered quantized hall systems.
\newblock {\em Phys. Rev. B}, 84:205327, 2011.

\bibitem{maciejko_prl_2010}
Joseph Maciejko, Xiao-Liang Qi, H.~Dennis Drew, and Shou-Cheng Zhang.
\newblock Topological quantization in units of the fine structure constant.
\newblock {\em Phys. Rev. Lett.}, 105:166803, Oct 2010.

\bibitem{tkachov_prb_2011}
G.~Tkachov and E.~M. Hankiewicz.
\newblock Anomalous galvanomagnetism, cyclotron resonance, and microwave
  spectroscopy of topological insulators.
\newblock {\em Phys. Rev. B}, 84:035405, Jul 2011.

\bibitem{brune_prl_2011}
C.~Br\"une, C.~X. Liu, E.~G. Novik, E.~M. Hankiewicz, H.~Buhmann, Y.~L. Chen,
  X.~L. Qi, Z.~X. Shen, S.~C. Zhang, and L.~W. Molenkamp.
\newblock Quantum hall effect from the topological surface states of strained
  bulk hgte.
\newblock {\em Phys. Rev. Lett.}, 106:126803, Mar 2011.

\bibitem{hancock_prl_2011}
Jason~N. Hancock, J.~L.~M. van Mechelen, Alexey~B. Kuzmenko, Dirk van~der
  Marel, Christoph Br\"une, Elena~G. Novik, Georgy~V. Astakhov, Hartmut
  Buhmann, and Laurens~W. Molenkamp.
\newblock Surface state charge dynamics of a high-mobility three-dimensional
  topological insulator.
\newblock {\em Phys. Rev. Lett.}, 107:136803, Sep 2011.

\bibitem{liu_prb_2010}
C.-X. Liu, X.-L. Qi, Zhang~H. J., Xi~Dai, Z.~Fang, and S.-C. Zhang.
\newblock Model hamiltonian for topological insulators.
\newblock {\em Phys. Rev. B}, 82:045122, 2010.

\bibitem{becker_pss_2007}
C.~R. Becker, C.~Br\"{u}ne, M.~Sch\"{a}fer, A.~Roth, H.~Buhmann, and L.~W.
  Molenkamp.
\newblock The influence of interfaces and the modulation doping technique on
  the magneto-transport properties of hgte based quantum wells.
\newblock {\em physica status solidi (c)}, 4(9):3382--3389, 2007.

\bibitem{volkov_infrared_1985}
A.~A. Volkov, Yu.~G. Goncharov, G.~V. Kozlov, S.~P. Lebedev, and A.~M.
  Prokhorov.
\newblock Dielectric measurements in the submillimeter wavelength region.
\newblock {\em Infrared Phys.}, 25(1-2):369, 1985.

\bibitem{pimenov_prb_2005}
A.~Pimenov, S.~Tachos, T.~Rudolf, A.~Loidl, D.~Schrupp, M.~Sing, R.~Claessen,
  and V.~A.~M. Brabers.
\newblock Terahertz conductivity at the verwey transition in magnetite.
\newblock {\em Phys. Rev. B}, 72(3):035131, Jul 2005.

\bibitem{berreman_josa_1972}
D.~W. Berreman.
\newblock Optics in stratified and anisotropic media - 4x4-matrix formulation.
\newblock {\em J. Opt. Soc. Am.}, 62(4):502, 1972.

\bibitem{shuvaev_epjb_2011}
A.~M. Shuvaev, S.~Engelbrecht, M.~Wunderlich, A.~Schneider, and A.~Pimenov.
\newblock Strong dynamic magnetoelectric coupling in metamaterial.
\newblock {\em Eur. Phys. J. B}, 79:163--167, 2011.
\newblock 10.1140/epjb/e2010-10493-1.

\bibitem{shuvaev_prl_2011}
A.~M. Shuvaev, G.~V. Astakhov, A.~Pimenov, C.~Br\"une, H.~Buhmann, and L.~W.
  Molenkamp.
\newblock Giant magneto-optical faraday effect in hgte thin films in the
  terahertz spectral range.
\newblock {\em Phys. Rev. Lett.}, 106:107404, Mar 2011.

\bibitem{hols_prl_2002}
F.~Hohls, U.~Zeitler, R.~J. Haug, R.~Meisels, K.~Dybko, and F.~Kuchar.
\newblock Dynamical scaling of the quantum hall plateau transition.
\newblock {\em Phys. Rev. Lett.}, 89:276801, Dec 2002.

\bibitem{ikebe_prl_2010}
Y.~Ikebe, T.~Morimoto, R.~Masutomi, T.~Okamoto, H.~Aoki, and R.~Shimano.
\newblock Optical hall effect in the integer quantum hall regime.
\newblock {\em Phys. Rev. Lett.}, 104:256802, Jun 2010.

\bibitem{sondi_rmp_1997}
S.~L. Sondhi, S.~M. Girvin, J.~P. Carini, and D.~Shahar.
\newblock Continuous quantum phase transitions.
\newblock {\em Rev. Mod. Phys.}, 69:315--333, Jan 1997.

\bibitem{novoselov_nature_2005}
K.~S. Novoselov, A.~K. Geim, S.~V. Morozov, D.~Jiang, M.~I. Katsnelson, I.~V.
  Grigorieva, S.~V. Dubonos, and A.~A. Firsov.
\newblock Two-dimensional gas of massless dirac fermions in graphene.
\newblock {\em Nature (London)}, 438:197, 2005.

\bibitem{buttner_nphys_2011}
B.~B\"{u}ttner, C.~X. Liu, G.~Tkachov, E.~G. Novik, C.~Br\"{u}ne, H.~Buhmann,
  E.~M. Hankiewicz, P.~Recher, B.~Trauzettel, S.~C. Zhang, and L.~W. Molenkamp.
\newblock Single valley dirac fermions in zero-gap hgte quantum wells.
\newblock {\em Nature Physics}, 7(5):418--422, MAY 2011.

\end{thebibliography}

\end{document}